\documentclass[aps, letterpaper, nofootinbib, showpacs, showkeys]{revtex4}

\usepackage{amssymb,amsmath,xcolor,graphicx,xspace,colortbl,textcomp, rotating} 
\usepackage{amssymb,amsmath,latexsym,xcolor,graphicx,xspace,colortbl,textcomp, rotating}  
\usepackage{varioref}  
\usepackage[dvips,dvipdfmx,hypertex]{hyperref}  
\graphicspath{{axisym727a_graphics/}{axisym727a_tcache/}{axisym727a_gcache/}}
\DeclareGraphicsExtensions{.pdf,.eps,.ps,.png,.jpg,.jpeg}
\DeclareGraphicsExtensions{.pdf,.eps,.ps,.png,.jpg,.jpeg}

\begin{document}
\title{On a New Formulation of Vacuum Axisymmetric Solutions of General Relativity }
\author{Louis Witten}
\email[E-mail me at: ]{lwittenw@gmail.com }
\affiliation{University of Florida }
\date{
\today
}
\begin{abstract}Axisymmetric solutions of the vacuum Einstein equations are found in the Papapetrou-Weyl gauge. The solutions depend on two pairs of
functionals, each pair of two functions depends on a different arbitrarily chosen function of one variable. Some examples are given. 
\end{abstract}
\pacs{04.20.Jb }
\keywords{axisymmetric solutions }
\maketitle

\section{Static Axisymmetric Solutions}
\qquad A standard and long known technique exists for finding static axisymmetric solutions to Einstein's theory in a Weyl
coordinate system. The metric in this coordinate system is
\begin{equation} ds^{2} =e^{2 \lambda }  dt^{2} -e^{2 \left (\mu  -\lambda \right )} \left (d \rho ^{2} + dz^{2}\right ) -\rho ^{2} e^{ -2 \lambda } d \varphi ^{2} \label{1}
\end{equation}  The Einstein equations reduce to the following set:
\begin{align}\lambda _{\rho  \rho } +\frac{\lambda _{\rho }}{\rho } +\lambda _{z z} =0 \\
\mu _{\rho } =\rho  (\lambda _{\rho }^{2} -\lambda _{z}^{2}) \\
\mu _{z} =2 \rho  (\lambda _{\rho } \lambda _{z})\end{align}  It is recognized that equation (2) is the three dimensional Laplace equation in cylindrical coordinates for a function
with axisymmetric symmetry. The method often applied in the literature is to define a source and then solve the equation using the Green function. After
choosing a $\lambda $ the equations for $\mu $ are to be solved by integration. 

\section{ Stationary Axisymmetric Solutions}
There is a well developed method to generate a stationary axisymmetric solution from a static one. This has been reviewed in a recent paper \cite{A}.
The result of applying the technique is that, after solving equations (2-4), one must solve another pair of equations to determine the value of the rotational
velocity which is determined by a function, $\omega $. The equations are
\begin{align}4 \alpha  \rho  \frac{ \partial \lambda }{ \partial \rho } =\frac{ \partial \omega }{ \partial z} \label{5} \\
4 \alpha  \rho  \frac{ \partial \lambda }{ \partial z} = -\frac{ \partial \omega }{ \partial \rho } \label{6}\end{align}  Solving the Laplace equation (2) for $\lambda $ and then calculating $\mu $ from equations (3,4) following with $\omega $ determined by equations (5,6) yields a stationary axisymmetric solution of the Einstein equations with the metric
\begin{equation} ds^{2} =\frac{e^{2 \lambda }}{1 +\alpha ^{2} e^{4 \lambda }} \left ( dt -\omega  d \varphi \right )^{2} -\frac{1 +\alpha ^{2} e^{4 \lambda }}{e^{2 \lambda }} [e^{2 \mu }(d \rho ^{2} + dz^{2}) +\rho ^{2} d \varphi ^{2}] \label{7}
\end{equation}  

\section{ An Alternate Method of Finding Stationary Axisymmetric Solutions}
An unpleasant feature of the process is that with the solution derived this way there is no easy way to tell, since the behavior of $g_{t t}$ is not simply related to $\lambda $, what any of the properties of the resultant solution will be until after the solution is examined. Since the function,~$\omega $, is directly related to the rotational velocity, one may wish to run the procedure backwards. Instead of finding~$\lambda $ by solving equation (2) and then solving equations (3-6) to get the metric; it is possible to find an $\omega $ and then solve equations (3-6) to get the metric. For equation (6) to have a solution for $\lambda $, the function $\omega $ must satisfy the following integrability condition
\begin{equation}\frac{ \partial }{ \partial z}(\frac{1}{\rho } \frac{ \partial \omega }{ \partial z}) = -\frac{ \partial }{ \partial \rho }(\frac{1}{\rho } \frac{ \partial \omega }{ \partial \rho })
\end{equation}  or, on rewriting,
\begin{equation}\frac{ \partial ^{2}\omega }{ \partial \rho ^{2}} -\frac{1}{\rho } \frac{ \partial \omega }{ \partial \rho } +\frac{ \partial ^{2}\omega }{ \partial z^{2}} =0
\end{equation}  Thus it is possible to find a stationary axisymmetric solution by solving the set of equations (2,3,4,5,6) or
the set (3,4,5,6,9). The general solution to the equations (2,5,6,9) can readily be found and are given in equations (10) and (11) below. The arbitrary
constants and arbitrary functions involved are to be chosen to fit the initial and boundary values of the physical situation.
\begin{align}\frac{\omega }{\rho } =4 \alpha _{1} m_{1} \int _{o}^{\infty }e^{ -k z} A (k) \mathcal{C}_{1}^{1} (k \varrho )  dk +4 \alpha _{2} m_{2} \int e^{ -k z} B (k) \mathcal{C}_{1}^{2} (k \rho )  dk \\
\lambda  = -m_{1} \int _{o}^{\infty }e^{ -k z} A (k) \mathcal{C}_{0}^{1} (k \rho )  dk -m_{2} \int _{0}^{k}e^{ -k z} B (k) \mathcal{C}_{0}^{2} (k \rho )  dk\end{align}  $\alpha _{1} ,\alpha _{2\;\text{,}\;} m_{1\;\text{,}\;} m_{2}$ are arbitrary constants, $A$ and $B$ are arbitrary functions, not necessarily continuous as long as the integrals exist. ~$\mathcal{C}_{1}^{1}$ and~$\mathcal{C}_{1}^{2}$ are any two functions of the set of functions $J_{1} (k \rho ) ,Y_{1} (k \rho ) ,H ,(k \rho ) ,H_{2} (k \rho )$, the Bessel and Hankel functions of the first order and first and second kind respectively. The functions with the $0$ subscript are the same functions but of the zeroth order. The functions chosen should be matched to the boundary conditions
of the problem. These choices assumed~$k$ to be real. It is equally possible to choose it to be purely imaginary, in which cases the two $\mathcal{C}$ functions chosen would be the two modified Bessel functions. The general solution to the equations giving~$\mu $ could be written in terms of a convolution of Bessel functions but it is usually more illuminating to keep it in terms of
equations (3,4). So the general solution depends on four functions and four constants. There are two pairs of functions, one pair depends on the a choice
of $A (k)$, $\alpha _{1}$, and $m_{1}$ and the other pair depends on the choice of $B (k)$, $\alpha _{2}$, and $m_{2}$. 

\section{Examples}
I shall give several examples of finding solutions to the vacuum equations. For simplicity I will choose~$B (k) =0$ and~$\mathcal{C}_{o ,1}^{1}$ to be Bessel functions of the first kind and zeroth and first order. The solutions are now written in somewhat modified form.
\begin{align}\frac{\omega }{\rho } =4 \alpha  m \int _{o}^{\infty }k e^{ -k z} \frac{A (k)}{k} J_{1} (k \rho )  dk \\
\lambda  = -m \int _{o}^{\infty }k e^{ -k z} \frac{A (k)}{k} J_{0} (k \rho )  dk\end{align}  Recall the definition of the Hankel transform of order $\nu $
\begin{equation}F_{\nu } (k) =\mathcal{H}_{\nu } (f (\rho ) ,k) =\int _{0}^{\infty }\rho  f (\rho ) J_{\nu } (k \rho )  d\rho 
\end{equation}  and its inverse
\begin{equation}f (\rho ) =\mathcal{H}_{\nu } (F_{\nu } (k) ,\rho ) =\int _{0}^{\infty }k F_{\nu } (k) J_{\nu } (k \rho )  dk
\end{equation}  The solutions (12) and (13) may be written
\begin{align}\frac{\omega }{\rho } =4 \alpha  m \mathcal{H}_{1} (\frac{e^{ -k z}}{k} A (k) ,\rho ) \\
\lambda  = -m \mathcal{H}_{0} (\frac{e^{ -k z}}{k} A (k) ,\rho )\end{align}  Choosing an arbitrary function $A (k)$ will yield a solution. Or choosing an $\omega /\rho $ whose Hankel transform goes like $A (k) \exp ^{ -k z}$ will yield a solution. There are tables of Hankel transform pairs and a choice of a pair that has the exponential as one of its functions
will give a solution. It is equally possible to consider equations (12) and (13) to be Laplace transforms of $A (k) J_{1} (k)$ and~$A (k) J_{0} (k)$ respectively and matching these with their Laplace transform twins. It may be more convenient, depending on boundary conditions to
use an imaginary $k$ together with modified Bessel functions and to match Fourier transforms. 

As a first example of a solution,
let~$A (k) =1$ and from a table of transforms of order $1$ and $0$ discover the solutions
\begin{align}\omega  =4 \alpha  m (1 -\frac{z}{(\rho ^{2} +z^{2})^{1/2}}) \\
\lambda  = -\frac{m}{(\rho ^{2} +z^{2})^{1/2}}\end{align}  and calculate
\begin{equation}\mu  = -\frac{1}{2} \frac{m^{2} \rho }{(\rho ^{2} +z^{2})^{2}}
\end{equation}  This is the solution generated from the Curzon solution. A less well known solution emerges if $A (k) =k$. In this case
\begin{align}\omega  =4 \alpha  m \frac{\rho ^{2}}{(\rho ^{2} +z^{2})^{3/2}} \\
\lambda  = -\frac{m z}{(\rho ^{2} +z^{2})^{3/2}}\end{align}  To calculate $\mu $, one may use equations (3,4) or equivalently
\begin{align}\mu _{\rho } =\frac{1}{16 \alpha ^{2} \rho } (\omega _{z}^{2} -\omega _{\rho }^{2}) \\
\mu _{z} = -\frac{1}{8 \alpha ^{2} \rho }(\omega _{z} \omega _{\rho )}\end{align}  Many solutions can be found from known solutions by using properties of Hankel transforms. For example, if $\lambda $ and $\omega /\rho $ are a solution, so are $ \partial ^{m}\lambda / \partial z^{m}$ and $ \partial ^{m}(w/\rho )/ \partial z^{m}$ for any $m$. The solution described by equations (21) and (22) seems to have assymptotically the same angular momentum as does the Kerr
solution. 

As a next example of a solution, I shall consider~$A (k) =\delta  (k -k_{0})$ and find
\begin{align}\omega  =4 \alpha  m \rho  e^{ -k z} J_{1} (k \rho ) \\
\lambda  = -m e^{ -k z} J_{0} (k \varrho ) \\
\mu  = -m^{2} \rho  k J_{0} (k \rho ) J_{1}(k \rho \end{align}  As a final example, I will give a complicated solution. The Kerr solution is also complicated in this coordinate system.
Let $A (k) =k J_{0} (a k)$ for constant $a$. This will introduce another parameter. In reference \cite{B},
the integrations to find $\omega /\rho $ and $\lambda $ appear as Laplace transforms of a Bessel function.
\begin{align}\omega  =\frac{2\alpha m\kappa  \rho ^{1/2 }}{ \pi  a^{3/2}} \left [\frac{\varkappa ^{2} \left (a^{2} -\rho ^{2} -z^{2}\right )}{4 \left (1 -\kappa ^{2}\right ) a \rho } \boldsymbol{E} \left (\kappa \right ) +\boldsymbol{K} \left (\kappa \right )\right ] \\
\lambda  = -m\frac{z \kappa ^{3}}{4 \pi  \left (1 -\kappa ^{2}\right ) \left (a \rho \right )^{3/2}} \boldsymbol{E} \left (\kappa \right )\end{align}  Three new functions have been introduced, $\kappa $, the elliptic integral of the first kind $\boldsymbol{K} (\kappa )$, and of the second kind, $\boldsymbol{E} \left (\kappa \right )$.
\begin{equation}\kappa  =\frac{2 \sqrt{a \rho }}{\sqrt{\left (a +\rho \right )^{2} +z^{2}}}
\end{equation}
\begin{equation}\boldsymbol{K} \left (\kappa \right ) =\int _{0}^{1}\frac{ dx}{\sqrt{1 -x^{2}} \sqrt{1 -\kappa ^{2} x^{2}}}
\end{equation}
\begin{equation}\boldsymbol{E} \left (\kappa \right ) =\int _{0}^{1}\frac{\sqrt{1 -\kappa ^{2} x^{2}}}{\sqrt{1 -x^{2}}}  dx
\end{equation}  

\section{ Conclusion and Comments}
To find a vacuum axisymmetric solution, one may choose $\omega $ to be any function of $\rho $ and $z$ whose first order Hankel transform exists and is of the form $\exp  ( -k z) A (k)$, or one may choose~$\exp  ( -k z) A (k)$ to be arbitrary. One may calculate either one from the other and then find $\lambda $. Or one may choose a $\lambda $ to satisfy equation (2) and then find its Hankel transform and from that calculate $\omega $. Several examples are given. The solution given by equations (21,22) may be of particular interest. The bridge between
$\lambda $ and~$\omega $ is given by equations (5,6). These equations are known and widely studied in hydrodynamics as being Stokes equation for
the streamlines in inviscid incompressible fluids.

\begin{acknowledgments}I thank D. B. Papadopoulos and K. Kleidis for discussions on this issue. K. Kleidis calculated equation 27.
\end{acknowledgments}

\end{document}